\documentclass[aps,prl,12pt,onecolumn, superscriptaddress]{revtex4-1}
\usepackage{amssymb,graphicx,color,amsmath,xfrac,bm,stmaryrd,trimclip}
\usepackage[mathletters]{ucs}
\usepackage[utf8x]{inputenc}
\usepackage{graphicx}%

\usepackage{multirow}%

\usepackage{amsmath,amssymb,amsfonts}%

\usepackage{amsthm}%

\usepackage{amsmath}%

\usepackage{mathrsfs}%

\usepackage[title]{appendix}%

 \usepackage{xcolor}%

\usepackage{textcomp}%

\usepackage{booktabs}%

\usepackage{algorithm}%

\usepackage{algorithmicx}%

\usepackage{algpseudocode}%

\usepackage{listings}%

\begin{document}

\title{Observation of the Yamaji effect in a cuprate superconductor}

\author{Mun K.~Chan}

\email[email: ]{mkchan@lanl.gov}

\affiliation{Pulsed Field Facility, National High Magnetic Field Laboratory,~Los Alamos National Laboratory,~Los Alamos,~New Mexico 87545, USA}

\author{Katherine A.~Schreiber}

\affiliation{Pulsed Field Facility, National High Magnetic Field Laboratory,~Los Alamos National Laboratory,~Los Alamos,~New Mexico 87545, USA}

\author{Oscar E. Ayala-Valenzuela}

\affiliation{Pulsed Field Facility, National High Magnetic Field Laboratory,~Los Alamos National Laboratory,~Los Alamos,~New Mexico 87545, USA}

\author{Eric D. Bauer}

\affiliation{Los Alamos National Laboratory, Los Alamos, New Mexico, 87545, USA}

\author{Arkady Shekhter}

\affiliation{Pulsed Field Facility, National High Magnetic Field Laboratory,~Los Alamos National Laboratory,~Los Alamos,~New Mexico 87545, USA}

\author{Neil Harrison}

\affiliation{Pulsed Field Facility, National High Magnetic Field Laboratory,~Los Alamos National Laboratory,~Los Alamos,~New Mexico 87545, USA}

\begin{abstract}
The pseudogap state of high-$T_{\rm c}$ cuprates, known for its partial gapping of the Fermi surface above the superconducting transition temperature $T_{\rm c}$~\cite{timusk99}, is believed to hold the key to understanding the origin of Planckian relaxation and quantum criticality~\cite{varma89,keimer15,phillips22,patel24}. However, the nature of the Fermi surface in the pseudogap state has remained a fundamental open question~\cite{norman1998,sebastian47,reber12,proust19,sobota21}. Here, we report the observation of the Yamaji effect~\cite{yamaji89,kartsovnik04,singleton2000} above $T_{\rm c}$ in the single layer cuprate HgBa$_2$CuO$_{4+\delta}$. This observation is direct evidence of closed Fermi surface pockets in the normal state of the pseudogap phase. The small size of the pockets determined from the Yamaji effect (occupying approximately $1.3\%$ of the Brillouin zone area) is all the more surprising given the absence of evidence for long-range broken translational symmetry that can reconstruct the Fermi-surface.
\end{abstract}
\date{\today}    
\maketitle    

\section*{Introduction}

Although evidence for symmetry breaking has been found in the pseudogap phase~\cite{xia08,shekhter13, sato17,zhao17,bourges21,murayama19}, its  connection to quantum criticality has not yet been established. For such symmetry breaking to be relevant for quantum criticality, it must also have a direct effect on the Fermi-surface~\cite{sebastian47,proust19}. Indeed, Fermi surface reconstruction has been explored in the cuprates through magnetic quantum oscillation and angle-dependent magnetoresistance measurements~\cite{leyraud07,sebastian14,ramshaw17,kunisada20}. However, these measurements have thus far been conducted either in a magnetic field-induced charge-density wave phase~\cite{gerber15,chang16} that exists at temperatures well below $T_{\rm c}$~\cite{leboeuf07,chan20}, or in an antiferromagnetic phase present at very low hole dopings ~\cite{kunisada20}. Because these phases are restricted to narrow low-temperature or low-doping regimes, they are not indicative of the Fermi surface in the broader underdoped region and high-temperatures spanned by the pseudogap.

\begin{figure}[t!h!]
\centering
\includegraphics[width=1.0\textwidth]{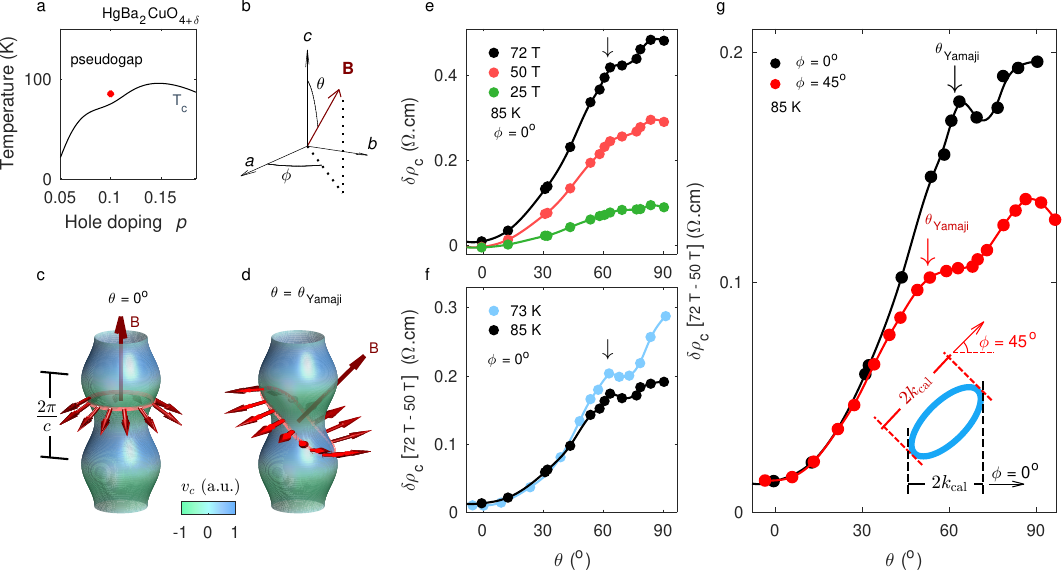}%
\caption{{\bf Observation of the Yamaji effect.} {\bf a}, Phase diagram of superconducting temperature versus doping of Hg1201 ~\cite{yamamoto00}. Red dot marks the focus of this work.  {\bf b,} Schematic of the polar ($\theta$) and azimuthal ($\phi$) angles. $a,b$ and $c$ are crystallographic directions. {\bf c}, Schematic example orbit on a quasi-2D Fermi-surface for $B\|c$. Arrows indicate the instantaneous velocity on the orbit. The $c$-axis component of the velocity, $v_c$, does not change on the orbit. {\bf d}, Schematic of example orbit for $\theta=\theta_{\rm Yamaji}$. $v_c$ oscillates around the orbit and averages to zero. {\bf e}, Polar magnetic field orientation-dependent curves of the magnetoresistivity $\delta\rho_c$ at three different fields, and at 85~K and $\phi=$~0$^\circ$. The arrow highlights the Yamaji peak, which disappears into the background at lower $B$ due to the lower $\omega_{\rm c}\tau$. {\bf f}, The same for two different temperatures. The peak is more pronounced at lower temperature due to a higher $\omega_{\rm c}\tau$. To emphasize the Yamaji peak and to avoid the effect of superconductivity at lower temperature, we plot the difference $\rho_{c}(72~$T$) - \rho_{c}(50~$T$)$. {\bf g}, Azimuthal angle-dependence of the Yamaji peak at $\phi=$~0$^\circ$ and 45$^\circ$. The inset schematically shows the relationship between caliper radius $k_{\rm cal}(\phi)$ and $\phi$ for an elliptical pocket. The caliper radius is determined directly from the Yamaji angle by $k_{\rm cal}(\phi) = 3\pi/4c \tan[\theta_{\rm Yamaji}(\phi)]$ where $c$ is the $c-$axis lattice parameter (see Methods)~\cite{kartsovnik04,singleton2000,grigoriev10}. We find $k_{\rm cal}(\phi=0^\circ) = 0.12\pm 0.01$~\AA$^{-1}$ and $k_{\rm cal}(\phi=45^\circ) = 0.16\pm0.02$~\AA$^{-1}$, respectively.
}
\label{yamaji}
\end{figure}

The conspicuous observation of anomolous `Fermi-arcs', and the lack of experimental evidence for closed pockets in the pseudogap state above $T_{\rm c}$~\cite{norman1998,damascelli04,sobota21}, has inspired suggestions of an unconventional Fermi surface~\cite{moon11,rice11} or even the total absence thereof~\cite{reber12,norman07,sobota21}. A recent study in La$_{2-y-x}$Nd$_y$Sr$_x$CuO$_4$ (Nd-LSCO)~\cite{fang22} found tantalizing evidence of a change in the angle-dependent magnetoresistivity across the supposed pseudogap critical doping.  However, this work could not determine if the Fermi-surface in the pseudogap comprises closed pockets~\cite{fang22}, nor could it definitively differentiate between many Fermi-surface scenarios, including changes of the Fermi-surface due to a Lifshitz transition of the underlying electronic band structure~\cite{musser22}. This stems from the short lifetime of the carriers suppressing the Yamaji effect~\cite{yamaji89}, observation of which is essential for demonstrating closed Fermi surface pockets and for the direct determination of their size. Consequently, the Fermi surface footprint of the pseudogap above $T_{\rm c}$ relevant for quantum criticality has continued to remain an open question.

The Yamaji effect manifests as peaks in the interlayer, or $c$-axis, resistivity $\rho_c$, of a quasi two-dimensional (layered) metal when the orientation of the magnetic field is tilted away from the $c$-axis~\cite{singleton2000,kartsovnik04}. The criterion for the observation of a Yamaji peak is similar to that for observing magnetic quantum oscillations: the product $\omega_{\rm c}\tau$ of the cyclotron frequency $\omega_{\rm c} = eB/m^\ast$ (where $e$ is the electric charge, $B$ is the magnetic field and $m^\ast$ is the quasiparticle effective mass) and the transport relaxation time $\tau$ must be at least of the order of unity: {\it i.e.} a significant fraction of electrons must complete a cyclotron orbit without scattering. Therefore, similar to quantum oscillations, observation of the Yamaji effect constitutes direct experimental evidence for a closed Fermi surface~\cite{singleton2000,kartsovnik04}. Unlike quantum oscillations, which can only be observed at low temperatures when cyclotron orbits are quantized~\cite{shoenberg}, the Yamaji effect can be observed at much higher temperatures because it does not require orbital quantization~\cite{yagi90}. In this respect, the Yamaji effect is particularly suited for Fermi-surface studies at the higher temperatures necessary to understand the pseudogap phase from which the high-$T_c$ superconductivity emerges.

\section*{Results}
Figure~\ref{yamaji} shows measurements of the Yamaji effect in an underdoped cuprate HgBa$_2$CuO$_{4+\delta}$ (Hg1201) with a hole doping $p = 0.10$ above its superconducting temperature of $T_c = 74~$K. We concern ourselved here with the magnetic field induced change in $c$-axis resistivity, or magnetoresistivity, $\delta\rho_{c} = \rho_{c}(B)-\rho_{c}(0)$. In a static magnetic field, one can rotate the polar angle $\theta$ while recording $\delta\rho_{\rm c}$ \textit{in situ}~\cite{hussey03,fang22}. However, achieving high values of $\omega_{\rm c}\tau$ at  elevated temperatures in the cuprates requires the use of pulsed magnetic fields. High magnetic field curves of the magnetic field angle-dependent $\delta\rho_{c}$, such as those shown in Fig.~\ref{yamaji}e, are obtained from many individual high magnetic field sweeps (see Supplementary Figs.~\ref{thetaDepRaw},\ref{phiDepRaw}). 
The Yamaji effect is evident as a local maximum in $\delta\rho_c$ plotted against $\theta$, as indicated by the arrow in Fig.~\ref{yamaji}e. Figs.~\ref{yamaji}e,f show that the angular position of the peak, $\theta_{\rm Yamaji}$, does not shift with temperature or magnetic field. It does, however, shift with the azimuthal angle of the magnetic field $\phi$ (see Fig.~\ref{yamaji}g)~\cite{singleton2000,kartsovnik04}. The peak also becomes weaker and eventually fades into the background with decreasing $B$, as shown in Fig.~\ref{yamaji}e. 

The Yamaji effect results fundamentally from the $c$-axis hopping of electrons between layers, which causes warping (or corrugation) of what would otherwise be a cylindrical Fermi surface, as shown schematically in Figs.~\ref{yamaji}c,d~\cite{singleton2000,kartsovnik04}. The simple tetragonal structure of Hg1201~\cite{putilin93} leads to a particularly simple form of warping captured by the lowest harmonic sinusoidal. The effect of such warping on the conductivity is understood by examining the velocity evolution along classical trajectories on the Fermi surface under the action of the Lorentz force. For the $c$-axis conductivity, $\sigma_c$, it suffices to trace only the $c$-axis projection of the velocity on the Fermi-surface, $v_c$, along the quasiparticle trajectory. Specifically, $\sigma_c$ is proportional to the Fourier transform of the time-delayed velocity correlation, $\sigma_c(\omega) \propto \langle v_c(\omega) v_c(-\omega) \rangle = \int\limits_0^{\infty} dt \exp\{ i \omega  t \} \langle v_c(t) v_c(0) \rangle_{\mathbf{k}}$. The time-delayed velocity correlation $\langle v_c(t) v_c(0) \rangle_{\mathbf{k}}$, is averaged over all possible starting points $\mathbf{k}$ on the Fermi surface~\cite{singleton2000,kartsovnik04}. The resistivity $\rho_c$ is then obtained by inverting the conductivity $\sigma_c(\omega)$ and taking the zero-frequency limit. 

The more effectively $v_c$ averages to zero around a cyclotron orbit, the smaller is the $c$-axis conductivity,  and therefore the larger is the $c$-axis resistivity. Such a physical picture enables a qualitative understanding of the results in Fig.~\ref{yamaji}. For very long relaxation times, $\omega_{\rm c} \tau \gg 1$, electrons can traverse the cyclotron orbit multiple times and the c-axis resistivity approaches the `clean’ limit. In this limit, the conductivity corresponds to the average velocity of a single complete orbit. As the scattering becomes more intense, and  $\omega_{\rm c} \tau $ decreases, the time-delayed velocity correlation is suppressed at long times such that the magnetic field dependence of the $c$-axis resistivity weakens and eventually disappears in the `dirty limit', $\omega_{\rm c} \tau \ll1$

\begin{figure}[t]
\centering
\includegraphics[width=0.655\textwidth]{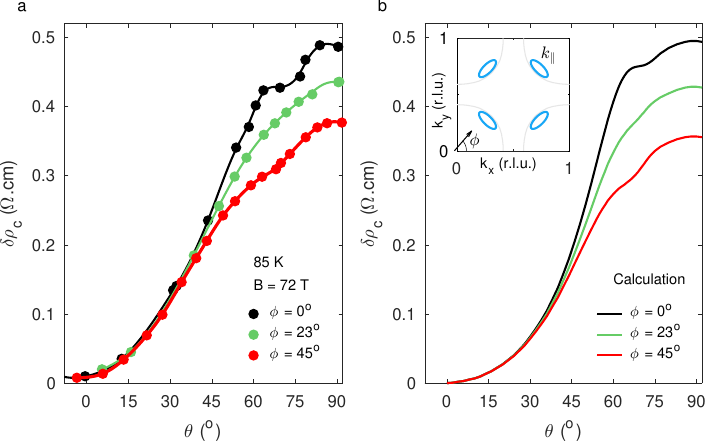}%
\caption{{\bf Comparison of measured angle dependent magnetoresistivity with a Boltzmann transport model}.
{\bf a}, Magnetoresistivity $\delta\rho_{c}$ at $72$~T and $85~$K as a function of $\theta$ for $\phi = 0^\circ$, $23^{\rm o}$ and $45^{\rm o}$. Full underlying data is shown in Supplementary Fig.~\ref{thetaDepRaw}. {\bf b}, Equivalent simulated $\delta\rho_{c}(\theta)$ (see Methods). Inset: Fermi-surface used for simulations.}
\label{model}
\end{figure}
When $B$ is aligned along the $c$-axis, the $c$-axis component of the velocity does not change around any given orbit for the simply warped Fermi-surface of Hg1201, ($\theta = 0^\circ$ in Fig.~\ref{yamaji}c). Hence, the conductivity is the same as that at $B=0$ such that $\delta\rho_c\approx 0$ at $\theta = 0^\circ$, as observed~(Fig.~\ref{yamaji}e). A very weak magnetoresistivity for $B$ along symmetry directions is well known in conventional metals with rotationally symmetric Fermi surfaces~\cite{abrikosov}. Upon rotating the magnetic field away from the $c$-axis, the cyclotron trajectories tilt, causing $v_c$ to not only change significantly around an orbit but also to change sign for an increasingly large fraction of the trajectory, yielding a smaller overall conductivity. This is reflected in the observed positive $\delta\rho_c$ at finite $\theta$ (in Fig.~\ref{yamaji}e). Upon reaching the Yamaji angle, $\theta_{\rm Yamaji}$, all cyclotron orbits have the same cross-sectional area~\cite{yamaji89,yagi90,kartsovnik02} and the averaged velocity for all {\it complete} orbits vanishes. As a result, the {\it c}-axis conductivity is fully suppressed in the clean limit, causing a large Yamaji peak in the polar angle dependence of the magnetoresisitivity. 

As a geometric consequence of a warped Fermi surface, the Yamaji angle should not shift with $\omega_{\rm c}\tau$. For small $\omega_{\rm c}\tau$, however, the Yamaji peak is attenuated and disappears into the background because relatively few electrons are able to complete cyclotron orbits. Hence, the conductivity no longer vanishes. This occurs, for example, when we decrease the magnetic field (Fig.~\ref{yamaji}e) or increase the temperature (Fig.~\ref{yamaji}f). At 72~T, $\omega_{\rm c}\tau \approx 2.6$ at 85~K (see Methods), which is sufficiently large to produce a clear Yamaji peak. This unambiguous evidence of  quasiparticles traversing complete orbits, implies a closed Fermi-surface pocket in the pseudogap state above $T_c$. One reason prior measurements on Nd-LSCO~\cite{fang22} failed to reveal a Yamaji effect was their significantly smaller values of $\omega_{\rm c}\tau$. For lower symmetry systems, such as Nd-LSCO~\cite{fang22} with staggered copper-oxide planes, the warping of the Fermi surface is also more complex. In such a case, the orbital average of $v_c$ is zero even at $\theta = 0^\circ$, leading a finite $c$-axis magnetoresistivity. The interpretation of the magnetoresistance in such lower-symmetry systems requires additional model assumptions~\cite{musser22,fang22,lewin18}. 

Our observation of the Yamaji effect also provides a direct measure of the in-plane size of the Fermi-surface. The Yamaji angle is directly related to the maximum in-plane Fermi momentum projection along the applied magnetic field, termed the `caliper' radius,  $k_{\rm cal}(\phi)$ (see {\it e.g.} schematic inset of Fig.~\ref{yamaji}g  and Methods)~\cite{kartsovnik04,singleton2000,grigoriev10}. For an elliptical pocket, the location of the Yamaji peaks $\theta_{\rm Yamaji}$ in Fig.~\ref{yamaji}g, implies a pocket area occupying approximately 1.3\% of the in-plane Brillouin zone  with a major axis radius that is approximately 2.6 times larger than the minor axis radius (see Methods). Regardless of the exact shape of the Fermi-pocket, our observation of the Yamaji peaks provides direct evidence of a small-closed Fermi-pocket in the  pseudogap state above $T_{\rm c}$.

\begin{figure}[t]
\centering
\includegraphics[width=0.655\textwidth]{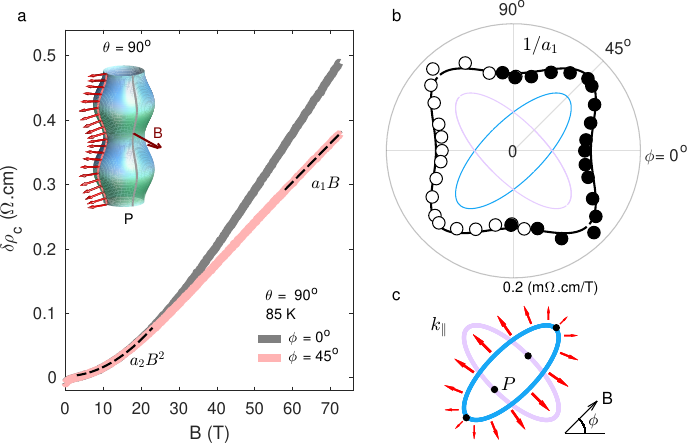}%
\caption{{\bf Magnetoresistivity for in-plane magnetic field.}
{\bf a}, $\delta\rho_c$ for $\theta =$ 90$^\circ$. $\delta\rho_c$ is linear at high field with slope $a_1(\phi)$, and quadratic at low field. The inset shows a schematic open orbit on the Fermi-pocket. On the trajectory marked $P$, the magnetic field  is parallel to the in-plane projection of the velocity, $\mathbf{v}_{\|}\parallel \mathbf{B}$ such that quasiparticles do not experience a Lorentz force. {\bf b}, Azimuthal angle-dependence of $1/a_1(\phi)$ (circles) extracted from fitting measured magnetoresistivity, see Supplementary Fig.~\ref{phiDepRaw} for underlying magnetoresistivity data. Open circles are mirrored from the measured data. The slope is related to the geometry of the quasi-2D Fermi surface as $1/a_1 (\phi)\propto R_\|(\phi)|/v_\|(\phi)^2$ (see Methods)~\cite{lebed97}.  $R_\|(\phi)$ and $v_\|(\phi)$ are the in-plane radius of curvature and velocity evaluated at points $P(\phi)$. The blue and purple lines are the $1/a_1 (\phi)$ of each of the two orthogonally oriented elliptical Fermi-surface cross-sections in our model. The solid black line is the sum of these two contributions, resulting in the fourfold planar anisotropy. {\bf c,} Planar elliptical cross-sections of the Fermi-pockets superimposed to demonstrate the resulting anisotropy of $1/a_1(\phi)$. Red arrows represent $v_\|$ around one of the pockets. Black dots indicate points $P$ on the Fermi-surface which determine $a_1$ for the example of $\phi=45^\circ$.
}
\label{fourfold}
\end{figure}

Our observed magnetoresistivity can also be compared against Boltzmann transport simulations~\cite{kartsovnik04, hussey03, fang22}. A simple model Fermi surface that captures our observations has a planar cross-section comprising two sets of orthogonally oriented ellipses, depicted in the inset of Fig.~\ref{model}b. Fig.~\ref{model} shows a side-by-side comparison of measurements of $\delta\rho_c$ and simulations at three different azimuthal angles. The size and shape of the pockets is constrained by the Yamaji effect observations such that the simulation employs only two adjustable parameters: the product of the cyclotron frequency and scattering time $\omega_{\rm c}\tau$, and the $c$-axis hopping amplitude $t_c$ (see Methods for values). Fig~\ref{model} shows that the observed angle-dependent magnetoresistivity, including the magnitude and location of Yamaji peaks, is captured by the simulation. 

A crucial observation is the strong azimuthal angle dependence of the Yamaji effect shown in Fig.~\ref{model}: a clear Yamaji peak for $\phi = 0^\circ$ (black curves), its disappearance at the intermediate angle $\phi = 23^\circ$ (green curves), and weak reappearance at $\phi=45^\circ$ (red curves). This can be understood as resulting from a superposition of magnetoconductivity of the two sets of `orthogonal' elliptical pockets. It is instructive to note that the Yamaji peak is, in fact, the first in an infinite series of peaks. While strongly damped at larger angles, these peaks are periodic in $\tan\theta$---referred to as angle-dependent magnetoresistance oscillations (AMROs)~\cite{singleton2000,kartsovnik04}. When two or more pockets are present, the total conductance is the sum of their respective contributions, and their AMROs `interfere' (illustrated in Supplementary Fig.~\ref{2pock}). The AMROs interfere constructively for the two sets of orthogonal pockets at $\phi=0^\circ$. At $\phi=45^\circ$, the `effective' $\omega_{\rm c}\tau$ is different at the Yamaji angles of the two pockets, causing the constructive interference to be incomplete and yielding a smaller Yamaji peak. By contrast, at the intermediate angle of $\phi=23^\circ$, the two sets of orthogonal pockets interfere destructively, rendering the Yamaji peaks indiscernible in our experiment. 

Further evidence for two sets of orthogonal pockets in Hg1201 can be found by considering the magnetoresistivity far from the Yamaji angle, when the magnetic field is oriented within the \(a,b\)-plane ($\theta = 90^\circ$), as shown in Figure~\ref{fourfold}a. Here, the Lorentz force gives rise to open orbits in momentum space as depicted schematically in the inset to Fig.~\ref{fourfold}a. The resultant magnetoresistivity is linear-in-field over a broad field range, and then crosses over to quadratic (i.e. $B^2$) field-dependence at lower fields~\cite{kartsovnik04} (see Methods). The high-field slope, $a_1(\phi)$, provides a further direct measure of the \(a,b\)-plane Fermi surface geometry~\cite{lebed97,smith10}. $a_1(\phi)$ exhibits a fourfold rotational symmetry in the azimuthal angle as shown in Fig.~\ref{fourfold}b (we have plotted the inverse of $a_1(\phi)$ for clarity). Were there only a single elliptical pocket, $a_1(\phi)$ would have two-fold rotational symmetry (blue and purple lines in Fig.~\ref{fourfold}b), contrary to observations. Although one pocket with four-fold rotational-symmetry in the center of the Brillouin zone can in-principle produce the observed symmetry of $a_1(\phi)$, it cannot yield the detailed evolution of the Yamaji effect with azimuthal angle, specifically the disappearance of the Yamaji peak at $\phi = 23^\circ$ (see  {\it e.g.} Supplementary Fig.~\ref{CDW}). The same Fermi-surface parameters used to simulate the Yamaji effect in Fig.~\ref{model} reproduces the measured $a_1(\phi)$ in Fig.~\ref{fourfold}b (solid lines, see Methods). All the magnetoresistivity data is captured by considering an isotropic $\tau$~\cite{grissonnanche2021}. 

\section*{Discussion}

The magnetoresistivity exhibiting a clear Yamaji peak, being a purely geometric Fermi-surface effect, can be understood  through the semiclassical evolution of the velocity under the Lorentz force acting on a {\it closed} Fermi surface. This establishes the existence of closed pockets in the {\it normal state} (above $T_c$) of Hg1201. The small size of the pockets is contrary to expectations of a large Fermi-surface from band structure calculations~\cite{das12}. It therefore raises the question of  its possible microscopic origin. So far, there is no evidence for conventional mechanisms producing pockets in Hg1201 above $T_{\rm c}$, such as antiferromagnetic or charge-density wave order, which involve translational symmetry-breaking~\cite{sebastian47,proust19}. Neutron scattering in Hg1201 in the pseudogap state finds only short-range antiferromagnetic fluctuations with a $\sim 30 $~meV spin gap~\cite{chan16,chan16c}. This rules out antiferromagnetic order. Reconstruction by short-ranged charge-density-wave or antiferromagnetic correlations present above $T_{\rm c}$~\cite{comin16,kampf94} is also precluded by the large value of \(\omega_{\rm c}\tau\). Using a direct relation between \(\omega_{\rm c}\tau\) and the mean free path $\lambda$, $\omega_{\rm c}\tau = \lambda/ \lambda_{\rm c}$ (where $\lambda_{\rm c}$ is the cyclotron radius~\cite{abrikosov}), the mean free path in Hg1201 at 85~K is $\lambda \approx 250$~\AA. 

Possible unconventional mechanisms for reconstructing the Fermi surface include those without broken translational symmetry (i.e., \(\mathbf{Q}=0\))~\cite{varma97,kivelson98,moon11,rice11}. While there have been experimental indications of \(\mathbf{Q}=0\) broken symmetries across the pseudogap temperature \(T^\ast\) in Hg1201~\cite{bourges21,murayama19}, their ability to reconstruct the Fermi surface remains uncertain. More broadly, the thermodynamic signatures associated with observed symmetry-breaking across \(T^\ast\)  of the cuprates~\cite{shekhter13,zhao17,keimer15} are too weak to account for Fermi surface reconstruction within the pseudogap region. The origins of the pseudogap phase of the cuprates must undoubtedly also explain the small-pockets observed here.


\bibliography{munpocket}

\clearpage
\newpage

\section*{Acknowledgements}

 \noindent We thank M. R. Norman for critical comments on a early version of the manuscript. The high-magnetic field measurements and sample preparation was supported by the US Department of Energy BES ‘Science of 100T’ grant. The National High Magnetic Field Laboratory - Pulsed-Field Facility is funded by the National Science Foundation Cooperative Agreement Number DMR-1644779, the State of Florida and the U.S. Department of Energy. MKC acknowledges support from LDRD-ER for calculations of electrical transport in unconventional superconductors. M.K.C. acknowledges support from NSF IR/D program while serving at the National Science Foundation. Any opinion, findings, and conclusions or recommendations expressed in this material are those of the author(s) and do not necessarily reflect the views of the National Science Foundation.

\section*{Author contributions statement}


\noindent M.K.C, K.S. ,O.E.A-V and N.H. developed equipment and performed pulsed field measurements. M.K.C. and E.D.B synthesized the samples. M.K.C, K.S. and N.H. analyzed and modeled the data. M.K.C., A.S. and N.H. interpreted the results and wrote the manuscript with critical input and review from all authors.

\section*{Competing Interests.}

\noindent The authors declare no competing interests.

\section*{Data Availability.}

\noindent Source and raw data underlying the results is available at  xxx.xxx


\newpage

\cleardoublepage
\section*{Supplementary Materials}
\section{Methods}

\noindent{\bf Sample preparation.} Hg1201 single crystals were grown using an encapsulated self-flux method~\cite{zhao06}. Samples from the same batch were used for a prior study showing quantum oscillations~\cite{chan20} at low temperatures. The measured sample is underdoped with $p = 0.1$ and $T_c = 74~$K. Newly synthesized crystals were heat treated at $450~$C$^\circ$ in flowing nitrogen for 1 month to achieve the desired hole doping. 

\hfill \break
\noindent{\bf Pulsed field measurements.} High magnetic field measurements were performed at the Pulsed-Field Facility of the National High Magnetic Field Laboratory at Los Alamos National Laboratory. Measurements were made with the 75 T duplex magnet whose pulse profile is shown in Supplementary Fig.~\ref{vxvy}c. Supplementary Fig.~\ref{vxvy}a,b demonstrates fidelity of the data collected during the magnetic field pulses.

\hfill \break
\noindent{\bf Calculation of resistivity.}
The conductivity of a metal is given by the Kubo formula:
 \begin{equation}
 \sigma_{\rm \alpha\beta}(\omega) = 
 \frac{1}{i\omega}
 \int\!\! dt\,e^{i\omega t}
 \,\langle\!\langle j_\alpha\, (t) j_\beta (0) \rangle\!\rangle\,,
\end{equation}
where $\langle\langle \dots \rangle\rangle$ are the thermodynamic and quantum averages. The static conductivity is obtained by taking the zero frequency limit $\omega\rightarrow 0$. In the semi-classical limit, the Kubo formula reduces to the average of the velocity over all possible classical trajectories:
 \begin{equation}\label{bs1}
 \sigma_{\rm \alpha\beta} = 2e^2 \frac{N_{\rm 2D}}{(2\pi)^2}\,\omega_{\rm c} \cos{\theta} 
 \!\int\limits_{-\frac{\pi}{c}}^{\frac{\pi}{c}}\frac{dk_0}{2\pi} 
 \!\!\int\!\! dt'
 \!\!\int\limits_0^{\infty}\!\!dt\,\, 
 v_{\alpha} (t+t^\prime)\, v_{\beta} (t^\prime)\,e^{-t/\tau} \,,
\end{equation}
where $N_{\rm 2D}$ is the average two-dimensional density of states at the chemical potential. The quasiparticle current is given by the product of electron charge and velocity $j_\alpha=ev_\alpha$. The factor of $2$ accounts for the two spin components. $\omega_{\rm c}=eB/m^\star$ is the cyclotron frequency. $m^\star$ is the in-plane effective mass, assuming an in-plane parabolic approximation to the dispersion. Integration over $t^\prime$ accounts for all possible starting points on a given trajectory. $k_0$  is the central $c$-axis momentum of each orbit, and its integration accounts for all orbits with different vertical offsets along  the $c$-axis direction of momentum space.  Meanwhile, $v(t)$ is the instantaneous velocity of a quasiparticle as its momentum-space location on the Fermi-surface evolves in time under a Lorentz force. The exponent accounts for scattering effects within the relaxation time approximation.

We first consider the case of a warped cylindrical Fermi surface ({\it i.e.} with a circular cross-section in the $ab$-plane). We then extend it to a Fermi-surface with an elliptical cross-section. The dispersion of a warped cylindrical Fermi-surface is
\begin{equation}
{\epsilon}({\bf k}) =\frac{ {\bf k}_\parallel^2}{2m^\star} -2\,t_c\cos(k_{\rm c}\,c)\,,
\label{eq:dispersion}
\end{equation}
where ${\bf k}_\parallel$ is the in-plane component of the momentum, $k_c$ is the $c$-axis component of the momentum, and $c$ is the $c$-axis lattice spacing. The second term represents the simplest form of warping along $c$, lowest-harmonic sinusoidal. The $c$-axis hopping $t_c$ is small, in the sense that mass $m_c$ associated with the the $c$-axis dispersion, $m_c = \hbar^2/(t_c\,c^2)$ is much larger than the in-plane mass, $m^\star/m_c\ll 1$. Such a weak $c$-axis hopping is responsible for the weak warping of the otherwise perfectly cylindrical Fermi-surface.

Equation~(\ref{bs1}) is more easily evaluated by converting the integration over time variables, $t$ and $t^\prime$, to azimuthal angles of the electron momentum, $\psi$ and $\psi^\prime$, on a closed orbit.  This change in variable is possible because $\psi$ is a monotonic function of $t$. Upon implementing the change in variable, we obtain:
\begin{equation}\label{eq:arkadymonster}
 \sigma_{\rm \alpha\beta} = 
2e^2 \frac{N_{\rm 2D}}{(2\pi)^2}\,\,\omega_{\rm c} \cos{\theta} \!\!
 \int\limits_{-\frac{\pi}{c}}^{\frac{\pi}{c}}\!\!dk_0\!\!
 \int_{0}^{2\pi}\!\!\! \frac{d\psi^\prime}{\Omega (\psi^\prime)} 
 \int\limits_{\psi'}^{\infty}\!\! \frac{d\psi}{\Omega (\psi)}  
 \,\,\,v_{\alpha} (\psi+\psi^\prime)\,v_{\beta} (\psi^\prime)\,
 \exp\left\{\!\!-\frac{1}{\tau}\!\!
 \int\limits_{\psi'}^{\psi}\!\! \frac{d\psi^{\prime\prime}}{\Omega (\psi^{\prime\prime})}\right\} \,,
\end{equation}
where $\Omega(\psi)$ is the Jacobian of the variable change from time $t$ to azimuthal angle $\psi$: $\Omega(\psi) = d\psi(t)/dt$. The Jacobian is $\Omega(\psi) = (d\psi/d\mathbf{k}_\parallel)\cdot (d\mathbf{k}_\parallel/dt)$, where $\mathbf{k}_\parallel$ is the in-plane projection of the electron momentum. The first factor is purely geometric, and connects the azimuthal angle $\psi$ to the in-plane momentum ${\bf k}_\|$. The second factor is equal to the force acting on the electron under classical evolution, $dk/dt = F$. For a Lorentz force, $F = e \mathbf{B}\times \mathbf{v}$, we obtain 
\begin{align}
\Omega(\psi) &= \frac{e}{k_{\parallel}^2}
\Big[ B_c\,\,\big(\mathbf{k_\parallel}\cdot\mathbf{v_\parallel}\big) -
v_c \,\,\big(\mathbf{k_\parallel}\cdot\mathbf{B_\parallel}\big) \Big]\,,    
\end{align}
where $k_{\parallel} =|\mathbf{k_\parallel}| $ and $\mathbf{v_\parallel}$ is the in-plane component of the  velocity. $B_c$ and $v_c$ are the $c$-axis components of magnetic field and instantaneous velocity respectively. The second term vanishes for zero warping because $v_c\rightarrow0$ when $t_c\rightarrow0$. At finite but small warping, the second term is smaller than the first by a small factor of order $m^\star/m_c$. To leading order of $m^\star/m_c$, we can ignore the second term such that the Jacobian depends only on the in-plane dispersion, as commonly done~\cite{yagi90,kartsovnik04,lewin18}. For a circular in-plane cross-section, Eq.~(\ref{bs1}), the Jacobian $\Omega(\psi)$ is independent of $\psi$ and is equal to $\Omega = \omega_{\rm c} \cos(\theta)$. 

For an elliptical pocket, the $c$-axis warping term of the dispersion, Eq.~(\ref{bs1}), is unchanged, whereas the planar term becomes 
\begin{align}
\epsilon_\parallel ({\bf k}_\parallel) = \frac{k_x^2}{2m_x}+\frac{k_y^2}{2m_y}\,,    
\end{align}
where $k_x$ and $k_y$ are components of the in-plane momentum along the major and minor axes of the elliptical pocket. For such an elliptical pocket, the Jacobian is given by 
\begin{align}
\Omega(\psi) &= \omega_{\rm c}\cos{\theta}\,\,\, \frac{m^\star\left(\frac{k_x^2}{m_x} + \frac{k_y^2}{m_y}\right)}{k_x^2+k_y^2}= 
\omega_{\rm c}\cos{\theta}\,\,m^\star\! \left(\frac{\cos^2\psi\,}{m_x} + \frac{\sin^2\psi}{m_y}\right)\,,
\end{align}
Also for an elliptical pocket, both the cyclotron mass $m^\star = (1/2\pi)\,dA/d\epsilon$, ($A$ is the area of the Fermi-surface) and the density of states mass $m^\star = dN/d\epsilon$, ($N$ is the total number of states enclosed by the Fermi-surface) are equal to each other, $m^\star = \sqrt{m_x m_y}$. With this, the two-dimensional density of states per spin component is $N_{\rm 2D} = m^*/2\pi\hbar^2$.

In this approximation, where the Jacobian is independent of $k_c$, the integral over $k_0$ in Equation (\ref{eq:arkadymonster}) can be done analytically~\cite{musser22}. To do so, we first point out that for an electron moving on a given orbit crossing the vertical axis at $k_0$, there is a well defined relation between its $c$-axis momentum component $k_c(\psi)$ and its planar momentum $\mathbf{k}_\parallel (\psi)$. Again, considering the limit of small $m^\star/m_c$, this relation is~\cite{kartsovnik04,yagi90}: 
\begin{align}\label{eq:kc2}
 k_c(\psi) &= k_{0}+K_{\rm H}(\psi)\tan{\theta}\, \qquad \text{where}\qquad
 K_{\rm H}(\psi)= k_{\parallel}(\psi)\cos(\psi - \phi)\,.
\end{align}
$K_{\rm H}(\psi)$ is the planar projection of the momentum $\mathbf{k}_\|$ on the azimuthal plane of the applied magnetic field at angle $\phi$. $\theta$ is the polar angle of the applied magnetic field.

As an electron traverses an orbit, $k_c(\psi)$ is a periodic function of $\psi$, between maximum value $k_{c,{\rm max}}$ and minimum value $k_{c,{\rm min}}$. Similarly, the extent of the orbit in the {\it ab}-plane is also finite, and is referred to as the caliper radius $k_{\rm cal}$:
\begin{align}
k_{\rm cal}=\frac{1}{2}\underset{\psi}{\{\max}-\underset{\psi}{\min\}}\, K_{\rm H}(\psi)\,.
\end{align}
It is easy to show~\cite{house96} that for an ellipse with semi-major and semi-minor axes $a$ and $b$ respectively rotated by an angle $\gamma$,
\begin{align}\label{eq:caliperellipse}
k_{\rm cal}=\sqrt{a^2\cos\{\phi-\gamma\}^2+b^2\sin\{\phi-\gamma\}^2}\,.
\end{align}

We now consider the $c$-axis conductivity starting from Eq.~(\ref{eq:arkadymonster}). To leading order in $m^\star/m_c$, the only dependence on $k_0$ in Eq.~(\ref{eq:arkadymonster}), comes from the dependence of $v_c(\psi)$ because it vanishes in the limit of zero warping. Specifically, $v_c (\psi)$ is entirely determined by the warping term in Eq.~(\ref{eq:dispersion}),
\begin{align}
    v_c(\psi) &= \frac{2\,c\,t_c}{\hbar}\,\, \sin[c\,k_c (\psi)]
            =\frac{2\,c\,t_c}{\hbar}\,\, \sin\!\Big[c\,k_{0}+c\,K_{\rm H}(\psi)\tan{\theta}\,\Big]\,,
\end{align}
where in the second equality sign, we substituted  Eq.~(\ref{eq:kc2}) to express $k_c(\psi)$ in terms of $k_\parallel(\psi)$.

Integrating the product of velocities in Eq.\ref{eq:arkadymonster}  over $k_0$,  
\begin{align}
    &\int \limits_{-\frac{\pi}{c}}^{\frac{\pi}{c}}\!\! dk_0\,\, v_c(\psi)\, v_c(\psi^\prime)\notag\\
    & = \left(\frac{2\,c\,t_c}{\hbar}\right)^2\,
    \int \limits_{-\frac{\pi}{c}}^{\frac{\pi}{c}}\!\!dk_0\,\, 
    \sin\!\Big(c\,k_0+c\,K_{\rm H}(\psi)\,\tan{\theta}\,\Big)\,
    \sin\!\Big(c\,k_0+c\,K_{\rm H}(\psi^\prime)\,\tan{\theta}\,\Big)\notag\\
    & = \frac{\pi}{c}\left(\frac{2\,c\,t_c}{\hbar}\right)^2 
    \cos\!\Big(\!\tan{\theta} \, \big[\,c\,K_{\rm H}(\psi)-c\,K_{\rm H}(\psi^\prime) \,\big]\Big)
\end{align}
Finally, we arrive at the following representation of $c$-axis conductivity
\begin{align}
 \sigma_c &= 
2e^2 \,\frac{N_{\rm 2D}}{(2\pi)^2}\,\,\omega_{\rm c} \cos(\theta)\, 
 \frac{\pi}{c}\left(\frac{2\,c\,t_c}{\hbar}\right)^{\!\!2} 
 \int\limits_0^{2\pi} \!\! \frac{d\psi^\prime}{\Omega (\psi^\prime)} 
 \int\limits_{\psi'}^{\infty}\!\! \frac{d\psi}{\Omega (\psi)} \,\,
  \exp\Big\{\!\!-\frac{1}{\tau}\!\!
 \int\limits_{\psi'}^{\psi}\!\! \frac{d\psi^{\prime\prime}}{\Omega (\psi^{\prime\prime})}\Big\} 
 \notag\\ 
&\qquad \qquad  \times 
    \cos\Big[\! \, \big(\,c\,K_{\rm H}(\psi)-c\,K_{\rm H}(\psi^\prime) \,\big)\tan{\theta}\Big] \,,
\end{align}
where $\,c\,K_{\rm H}(\psi)-c\,K_{\rm H}(\psi^\prime)$ is a periodic function of $\psi$ and $\psi'$ with maximum and minimum values of $\pm c\, 2k_{cal}$ respectively. In the clean limit, $\omega_{\rm c}\tau\gg 1$, the argument of the intergal is a periodic function of $\tan{\theta} \big[\,c\,K_{\rm H}(\psi)-c\,K_{\rm H}(\psi^\prime) \,\big]$. Therefore, upon integrating over $\psi$ and $\psi'$, the $c$-axis conductivity $\sigma_c$ is a periodic function of the amplitude of the expression under the cosine. In other words, the minima of $\sigma_c$ is periodic in  $c\, k_{\rm cal}(\phi)\tan{\theta} $. In particular the minima of this periodic function defines the set of Yamaji angles~\cite{yamaji89} 
\begin{align}\label{eq:caliper}
    c\, k_{\rm cal}(\phi)\tan[\theta_{\rm Yamaji}(\phi)]  = \frac{3}{4}\pi + n\pi \,\,\qquad n=0,\,1,\,2\dots \infty\,.
\end{align}
Although this expression for the Yamaji angle has been derived analytically for an elliptical pocket, it  has a pure geometric meaning. In particular, we remind the reader that $2k_{\rm cal}\tan{\theta}$ is the vertical extent of the orbit. The first Yamaji angle corresponds to when the vertical extent of the orbit is equal to half the distance from the belly of the first Brilluoin zone to the neck of the next zone~\cite{yamaji89}. Larger $\theta_{\rm Yamaji}$ corresponds to a smaller $k_{\rm cal}$.

The $c$-axis resistivity is determined by inverting the conductivity $\rho_{c} = 1/\sigma_{\rm c}$. For simulations, the conductivity contributions from each pocket is summed before inverting the total to obtain the resistivity. As described below, the Yamaji effect constraints the size and geometry of the Fermi-pockets. After assuming a simple Fermi-surface model comprising four ellipse cross-sections at each of the nodes, the only additional parameters for our simulations are $\omega_{\rm c}\tau$ and $t_c$. $\omega_{\rm c}\tau$ controls the size of the Yamaji peaks (Fig.~\ref{model}a,b) and the magnetoresistivity $\delta\rho_{c}$. The {\it c}-axes hopping $t_{c}$ controls the overall magnitude of the conductivity. The calculated conductivity is the sum of that from four pockets. 
 The resultant parameters for the model Fermi-surface are given in the table below.  

\begin{table}[ht]
\centering
\begin{tabular}{|c|c|c|c|}
\hline
$a$~(\AA$^{-1})$ & $a/b$ & $\omega_{\rm c}\tau$ & $t_{\rm c}$~(meV) \\
\hline
 $0.164\pm 0.03$ & $2.6$ & $2.6\pm 0.2$ &  $0.34\pm 0.02$  \\
\hline
\end{tabular}
\caption{\label{tab:example}Fermi-surface parameters. $a$ is the major axis radius of the elliptical cross-section of the Fermi-pocket. $a/b$ is the ratio between major and minor axis radii, or aspect ratio. }
\label{table1}
\end{table}

\noindent{\bf Fermi-pocket size and aspect ratio from the Yamaji effect.} The measured Yamaji angles determines the size and ellipticity of the Fermi-surface pockets {\it via} Eq.\ref{eq:caliper}. From Fig.~\ref{yamaji}g, we determine $\theta_{\rm Yamaji}(\phi = 0^{\rm o}) = 63.5\pm 2^{\rm o} $ and  $\theta_{\rm Yamaji}(\phi = 45^{\rm 0}) = 56.5\pm 3.5^{\rm o}$, corresponding to  $k_{\rm cal}(\phi=0^{\rm o}) = 0.124\pm 0.01$~\AA$^{-1}$ and  $k_{\rm cal}(\phi=45^{\rm o}) = 0.164\pm0.03$~\AA$^{-1}$. The latter corresponds to the major axis radius $a$, and the former permits a determination of the aspect ratio between major and minor axis radii $a/b =2.6$ using Eq.~\ref{eq:caliperellipse}. The enclosed area of each pocket is $A = \pi a b$ which comes out to be approximate 1.3\% of the area of the in-plane crystallographic Brillouin zone given by $(2\pi/3.9{\rm ~\AA})^2$. 

\hfill \break

\noindent{\bf Fermi-surface geometry from the linear-magnetoresistivity.} 
When the magnetic field is applied strictly within the $a$,$b$-plane ($\theta = 90^{\rm o}$), the quasiparticles follow open trajectories along $k_c$, as shown schematically in the inset of Fig.~\ref{fourfold}a. The quasiparticle trajectories are periodic in the reciprocal lattice unit $\delta k_c=2\pi/c$. The corresponding  {\it c}-axes cyclotron frequency is $\omega_{\perp}=(ec/\hbar)|\mathbf{v}_{\parallel}\times\mathbf{B}|$. $\mathbf{v}_{\parallel}$ is the planar component of the quasiparticle's instantaneous velocity~\cite{smith10,kartsovnik04}. 

The value of  $\omega_{\perp}$ for a particular trajectory depends on the magnitude of its planar velocity perpendicular to the applied magnetic field direction, as expected for motion driven by the Lorentz force. As such, $\omega_{\perp}$ can be large for some trajectories, but goes to zero at points on the Fermi-surface where $\mathbf{v}_{\parallel} \| \mathbf{B}$. The location of these points on the Fermi-surface depends on the azimuthal orientation of the applied magnetic field. An example, labeled $P$, is shown in Fig.~\ref{fourfold}c.  

Warping causes the interlayer velocity $v_c$ on these open trajectories to oscillate such that the average $v_c$ over a period is zero. This is the source of the large interlayer magnetoresistivity. For high magnetic fields,  trajectories with $\omega_{\perp}\tau\gg 1$ approach the `clean' limit such that their delayed velocity-velocity correlations are zero and thus do not contribute to the overall $\sigma_c$. However, $\omega_{\perp}\tau$ remains small for states on the Fermi-surface in the vicinity of $P$, whose planar velocity is almost parallel to the applied field. Such states dominate the conductivity at high fields, and are responsible for the observed linear-in-field magnetoresistivity. The number of such states, and thus the magnitude of magnetoresistivity, depends on the curvature of the Fermi-surface at $P$~\cite{kartsovnik04}.

It was shown in Ref.~\cite{lebed97} that the slope of the linear magnetoresistivity is obtained from the simple relation $a_1 (\phi)\propto v_\|(\phi)^2/R_\|(\phi)|$, where $R_\|(\phi)$ and $v_\|(\phi)$ are, respectively, the in-plane radius of curvature and the Fermi-velocity at the point $P(\phi)$ on the Fermi-surface. Therefore measurements of $a_1$ as a function of $\phi$ is a direct probe of the planar cross-sectional geometry of the quasi-2D Fermi-surface. For the case of the two sets of elliptical Fermi-pockets, the inverse slopes of each pocket can be simply summed (because $\sigma_c \sim 1/a_1$) to yield the total expected value. The result shows four-fold planar symmetry as depicted in Fig.~\ref{fourfold}b. 

\clearpage
\newpage
\section*{Supplementary Figures}

 \renewcommand\thefigure{S\arabic{figure}}    
 \setcounter{figure}{0}    


\begin{figure}[h]
\centering
\includegraphics[width=0.9\textwidth]{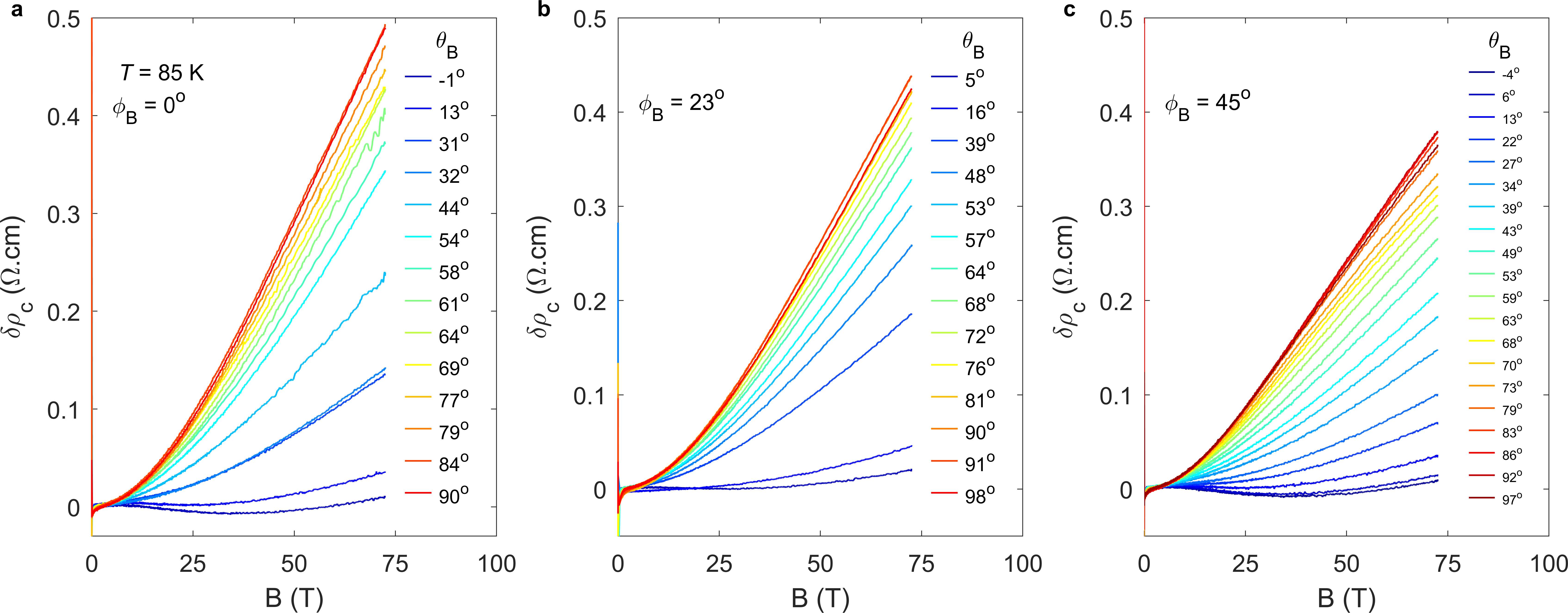}%
\caption{{\bf $\theta$ dependence of magnetoresistivity}. Field dependence of magnetoresistance underlying the $\theta$ dependence plots in Fig.~\ref{yamaji} and Fig.~\ref{model}. The magnetic field is tilted into $\phi = 0^{\rm o},~ 23^{\rm o}, {\rm and}~ 45^{\rm o}$ corresponding to each panel. $T = 85$~K.  
}
\label{thetaDepRaw}
\end{figure}

\clearpage

\begin{figure}[h]
\centering
\includegraphics[width=0.85\textwidth]{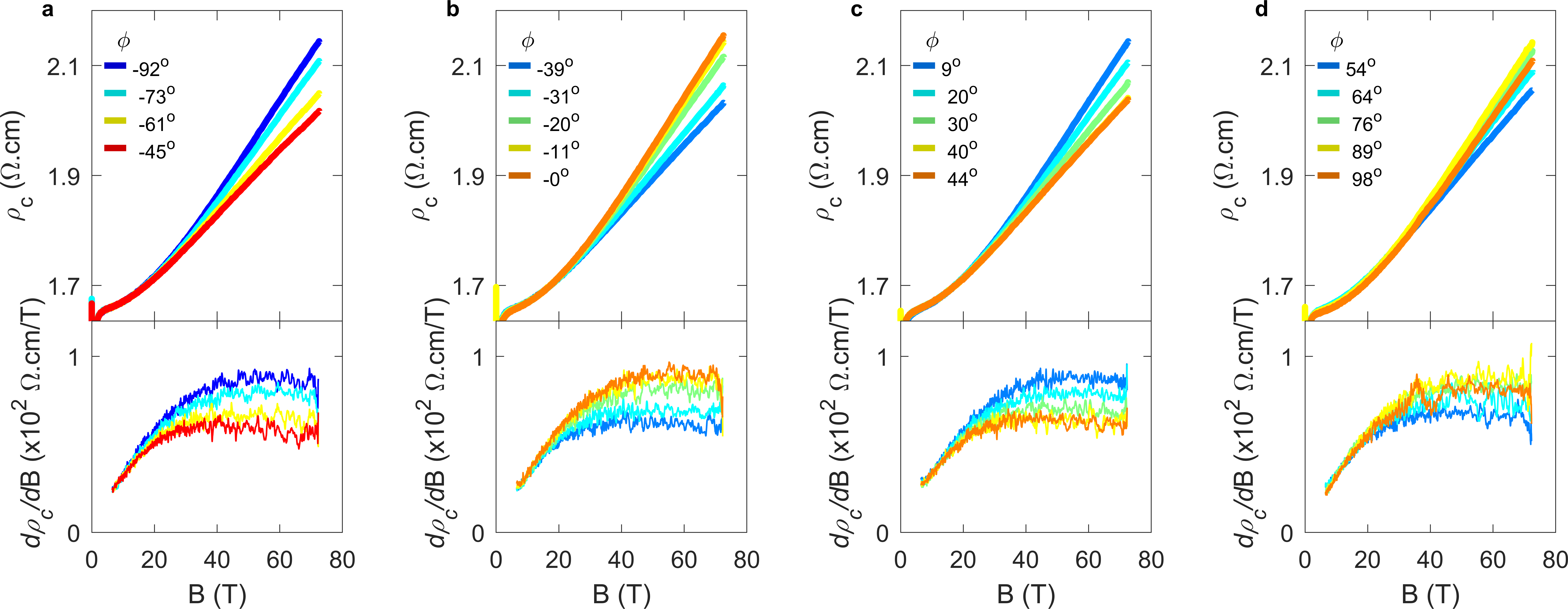}%
\caption{{\bf $\phi$ dependence of magnetoresistivity}. Field dependence of magnetoresistance underlying the $\phi$ dependence plots in Fig.~\ref{fourfold}b, covering $\approx 180^{\rm o}$ angular range. $\theta = 90^{\rm 0}$ and for all curves, corresponding to magnetic field applied in the plane. Lower panels are the derivative with respect to field, showing a crossover to linear magnetoresistance in all curves. $T = 85$~K.
}
\label{phiDepRaw}
\end{figure}

\clearpage

\begin{figure}[h]
\centering
\includegraphics[width=0.98\textwidth]{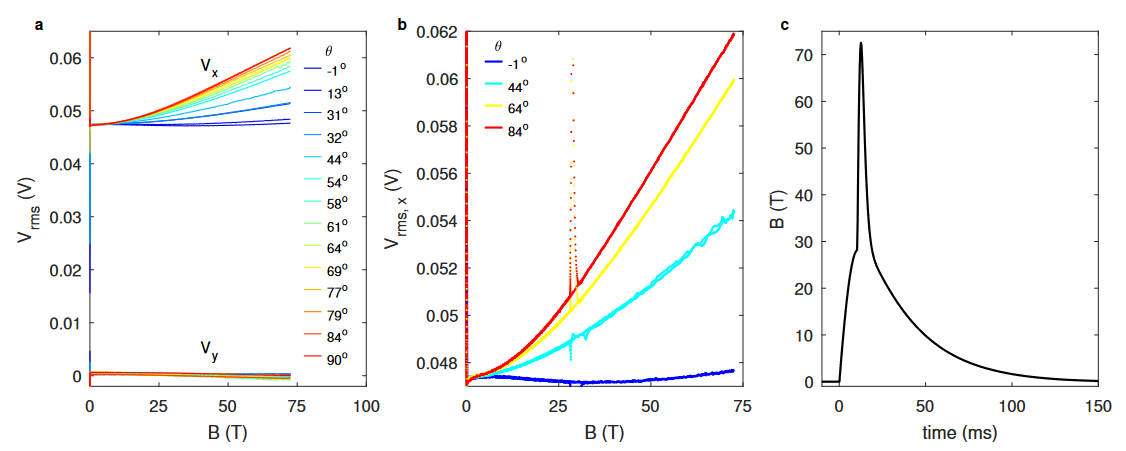}%
\caption{{\bf Fidelity of pulsed magnetic field measurements}. The resistance was measured via a four contact AC lockin technique at a frequency of 450 kHz. {\bf a,} Measured $V_x$ and $V_y$ of the lock-in voltage corresponding to $\theta$ dependence of magnetoresistance at $\phi = 0^{\rm o}$. The out-of-phase component $V_y$ is much smaller than the in-phase component $V_x$ of the detected voltage. {\bf b,} Representative field dependence of measured voltage comparing up (dots) and down (line) sweeps of the pulsed field. The two fall on top of each other except for a large spike starting around $\sim25$~T on the upsweep due to the firing of the insert.  {\bf c,} Pulsed field profile of the "75 Tesla Duplex" magnet pulsed to $\approx 72.5$~T
}
\label{vxvy}
\end{figure}

\clearpage

\begin{figure}[t]
\centering
\includegraphics[width=.95\textwidth]{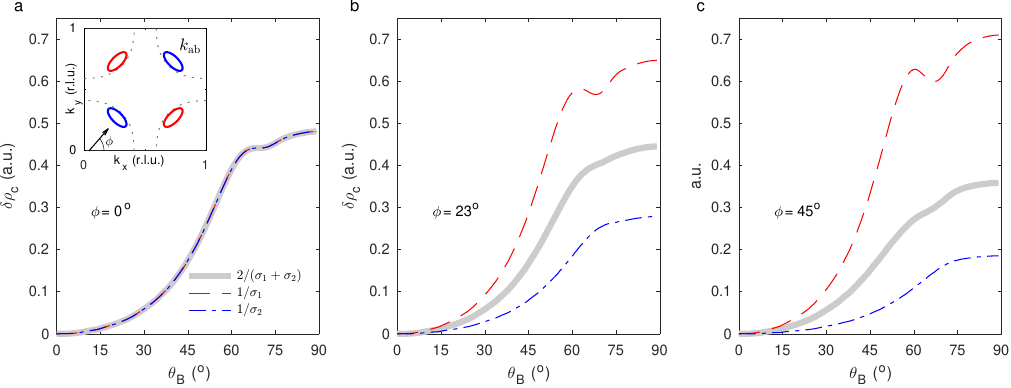}%
\caption{{\bf Contribution of two sets of orthogonal ellipses to the resistivity.}
{\bf a,} Simulated $\theta$ dependence of the magnetoresistivity $\delta\rho_c (\theta)$, with  $\omega_{\rm c}\tau = 2.6$, for each of the two orthogonally oriented ellipses  (red and blue, see inset for schematic of the Fermi-pockets color coded to match the curves) and the resultant combined contribution to $\delta\rho_c(\theta)$ of both sets of pockets (grey). The plotted combined $\delta\rho_c(\theta)$ curve includes a factor of two for easy comparison. For $\phi = 0^\circ$, $\delta\rho_c (\theta)$ is identical for the two sets of pockets. {\bf b,} For $\phi = 23^\circ$, $\delta\rho_c (\theta)$ of each ellipse are no longer the same. While the Yamaji peak of the red curve is clearly discernible, the Yamaji effect manifests only as a broad kink in the blue curve. The $\theta_{\rm Yamaji}$ of the two ellipse orientations are displaced such that the Yamaji peaks interfere destructively and is unobservable in the combined $\delta\rho_c (\theta)$ (grey). {\bf c,} For $\phi = 45^\circ$ the Yamaji peak of the red curve is sufficiently displaced from the kink in the blue curve such that the Yamaji effect is discernible as a small bump in the combined $\delta\rho_c (\theta)$ (grey).
}
\label{2pock}
\end{figure}

\clearpage

\begin{figure}[t]
\centering
\includegraphics[width=.7\textwidth]{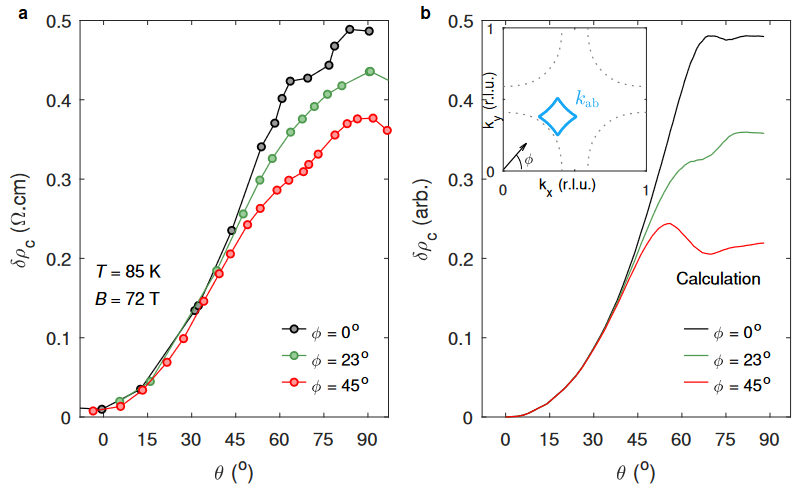}%
\caption{{\bf Comparing measurements of angle dependent magnetoresistivity to expectations for a bi-axial charge-density-wave reconstruction.}
{\bf a,} Magnetoresistivity of $\delta\rho_{c}$ as a function of magnetic field tilt from the {\it c}-axes into the plane along $\phi = 0^{\rm o}$, $23^{\rm o}$ and $45^{\rm o}$. {\bf b,} Simulations for a reconstructued Fermi-surface resulting from bi-axial charge-density-wave reconstruction previously studied at low temperatures ($T \lesssim 4$~K)~\cite{barisic13b,chan16b}. This pocket has four-fold planar symmetry in agreement with the symmetry of the measured $a_1(\phi)$, but it cannot capture the observed evolution of the Yamaji effect shown in panel {\bf a}.
}
\label{CDW}
\end{figure}



\clearpage

\begin{figure}[t]
\centering
\includegraphics[width=.95\textwidth]{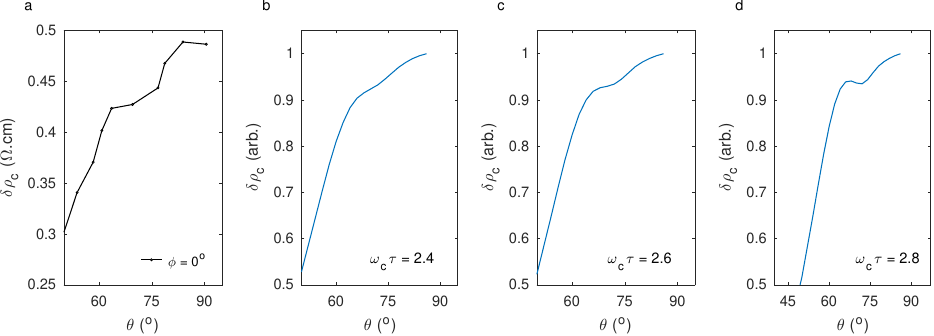}%
\caption{{\bf Effect of changing $\omega_{\rm c}\tau$ in modeling the Yamaji peak.}
{\bf a,} Close up of the Yamaji peak in $\delta\rho_{c}(\theta)$  at $\phi = 0^{\rm o}$. It is compared to calculations from our model with varying $\omega_{\rm c}\tau$ shown in panels {\bf b}-{\bf d}.
}
\label{tau}
\end{figure}

\clearpage

\end{document}